\documentclass[letterpaper, 10 pt, conference]{ieeeconf}  % Comment this line out
\IEEEoverridecommandlockouts                              % This command is only
\usepackage[utf8]{inputenc}
\usepackage[T1]{fontenc}
\usepackage{graphicx}
\usepackage{tablefootnote}

\title{\LARGE \bf
Machine Learning Method for Functional Assessment of Retinal Models
}

\author{Nikolas Papadopoulos$^{*}$\thanks{$*$ Nikolas Papadopoulos is with the School of Electrical and Computer Engineering, National Technical University of Athens, Greece {\tt \small nikpap555@gmail.com}},
Nikos Melanitis$^{\dag}$, Antonio Lozano$^{\ddag}$, Cristina  Soto-Sanchez$^{\ddag}$, Eduardo Fernandez$^{\ddag}$\thanks{$\ddag$ E.Fernandez,  C. Soto-Sanchez  and A. Lozano are with Instituto de Bioingenieria, Universidad Miguel Hernandez, Alicante, Spain}  \\and Konstantina S. Nikita$^{\dag}$\thanks{$\dag$ K.S. Nikita and N. Melanitis are with the School of Electrical and Computer Engineering, National Technical University of Athens, Greece}}

\usepackage{tikz}

\newcommand\copyrighttext{%
  \footnotesize \textit{© 2021 IEEE.  Personal use of this material is permitted.  Permission from IEEE must be obtained for all other uses, in any current or future media, including reprinting/republishing this material for advertising or promotional purposes, creating new collective works, for resale or redistribution to servers or lists, or reuse of any copyrighted component of this work in other works.}}
\newcommand\copyrightnotice{%
\begin{tikzpicture}[remember picture,overlay]
\node[anchor=north,yshift=-25pt] at (current page.north) {\fbox{\parbox{\dimexpr\textwidth-\fboxsep-\fboxrule\relax}{\copyrighttext}}};
\end{tikzpicture}%
}

\begin{document}

\maketitle

\copyrightnotice

\thispagestyle{empty}
\pagestyle{empty}

\vspace{-0.3cm}

%%%%%%%%%%%%%%%%%%%%%%%%%%%%%%%%%%%%%%%%%%%%%%%%%%%%%%%%%%%%%%%%%%%%%%%%%%%%%%%%
\begin{abstract}

Challenges in the field of retinal prostheses motivate the development of retinal models to accurately simulate Retinal Ganglion Cells (RGCs) responses. The goal of retinal prostheses is to enable blind individuals to solve complex, real-life visual tasks.  In this paper, we introduce the functional  assessment  (FA) of retinal models, which describes the concept of evaluating the performance of retinal models on visual understanding tasks. We present a machine learning method for FA:  we feed traditional machine learning classifiers  with  RGC responses  generated  by retinal models,  to solve object and digit recognition tasks (CIFAR-10, MNIST, Fashion MNIST, Imagenette). We examined critical FA aspects, including how the performance of FA depends on the task, how to optimally feed RGC responses to the classifiers and  how the number of output neurons correlates with the model’s accuracy. To  increase the number of output neurons, we manipulated input images - by splitting  and then feeding them to the retinal model - and we found that image splitting does not significantly improve the model's accuracy. We also show that differences in the structure of datasets result in largely divergent performance  of the retinal model (MNIST and Fashion MNIST exceeded 80\% accuracy, while \hbox{CIFAR-10} and Imagenette achieved $\sim$40\%). Furthermore,  retinal models which perform better in standard evaluation, i.e. more accurately predict RGC response, perform better in FA as well. However, unlike standard evaluation, FA results can be straightforwardly interpreted in the context of comparing the quality of visual perception.

% are closer to the biological retina, i.e. more accurately predict RGC response, perform better in FA. However, 

% we highlight the importance of FA, as FA results can be straightforwardly interpreted in the context (of quality) of (everyday) vision.

% We examined critical FA aspects: how the performance of FA depends on the task, how the number of output neurons correlates with the model’s accuracy, and how to optimally feed RGC responses to the classifiers. We show that differences in the structure of datasets result in largely divergent performance  of the retinal model (MNIST and Fashion MNIST exceeded 80\% accuracy, while CIFAR-10 and Imagenette achieved $\sim$40\%). Additionally,  retinal models that are closer to the biological retina, i.e. more accurately predict RGC response, perform better in FA. 

% However, FA results can straightforwardly be  interpreted in the context (of quality) of (everyday) vision?

% Finally, we found that retinal models who are closer to the biological retina perform better in FA.

% we suggest future directions towards the establishment of FA. 

\end{abstract}

\begin{keywords}
retinal models, functional assessment, machine learning, retinal prosthesis, visual recognition
\end{keywords}

%%%%%%%%%%%%%%%%%%%%%%%%%%%%%%%%%%%%%%%%%%%%%%%%%%%%%%%%%%%%%%%%%%%%%%%%%%%%%%%%
\section{INTRODUCTION}

The increasing knowledge of visual systems along with technological advances give novel results in prevention, limitation or even treatment of visual impairment \cite{lozano2018}, \cite{McIntosh2016}. However, retinal degenerative diseases, such as age-related macular degeneration (AMD) and retinitis pigmentosa cannot be effectively treated with surgery or medication \cite{fernandez2018}. Retinal prosthesis devices aim to restore vision in such patients, by translating visual stimuli to electrical stimulations that activate the retina. Then, the retina transmits neural signals to the visual centers of the brain, which are responsible for visual perception \cite{YueWeiland2016}. 

Although current retinal implants have managed to restore certain visual functionalities, there are several technological and biological challenges to be overcome. In particular, implants need to simulate natural retinal processing, by incorporating models that faithfully predict the physiological firing of retina cells to visual stimuli \cite{fernandez2018}, \cite{YueWeiland2016}. Towards this direction, the development of retinal models aims to simulate the biological neural processing in the retina, by interpreting images to retinal responses. Nowadays, advanced retinal models incorporate computer vision techniques \cite{Melanitis2019}, \cite{Alevizaki2019} and, more recently, state-of-the-art models use Convolutional Neural Networks (CNNs) \cite{lozano2018}, \cite{McIntosh2016}.

This paper proposes the functional assessment (FA) of retinal models, which describes the concept of evaluating the performance of retinal models on visual understanding tasks. This constitutes a divergence from the currently common practice of evaluating a retinal model comparing the similarity of model-generated and ground-truth RGC responses. Motivation for FA stems from the observation that visual prostheses aim to restore the capacity of individuals to comprehend their visual environment, thus we should directly evaluate our models on such tasks. The need for FA of prosthetic (i.e., acquired through prostheses) vision has been raised in the literature \cite{lepri2009}; FA has been applied to evaluate vision in implantees \cite{Demchinsky_2019} and in augmented-reality interventions in people with severe vision impairment \cite{angelopoulos2019}. FA should be focused on the visual functions that seem most important to the blind: mobility, face recognition and reading \cite{fernandez2018}. In this context,  we develop a machine learning method for FA. Using images from well-established computer vision datasets (Table \ref{table-datasets}), we feed traditional classifiers with RGC responses produced by the retinal model, in order to solve object and digit recognition tasks. Our goal is to explore whether retinal models that faithfully reproduce retina output show improved performance in visual understanding tasks and also, draw conclusions on the quality of prosthetic vision that is attainable by assessing the performance of retinal models directly on such tasks.

\section{METHODS}

\begin{figure}[htbp]
\centerline{\includegraphics[width=8cm]{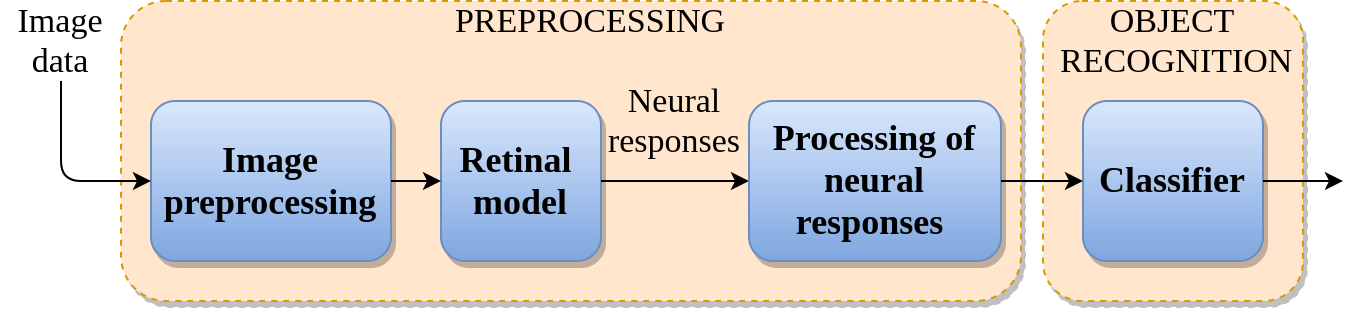}}
\caption{Functional assessment pipeline.}
% . Initially, we preprocess images (I) and give them as input to the retinal model (II). Then, we process the predicted RGC responses (III) and feed them to the classifiers for object and digit recognition tasks (IV).}
\label{method_pipeline}
\end{figure}

\subsection{Functional assessment pipeline} \label{section-data-preprocessing-steps}

Functional assessment pipeline (Fig. \ref{method_pipeline}) includes image preprocessing (Steps \ref{section-data_preprocessing-input}, \ref{section-data_preprocessing-data_augmentation}, \ref{section-data_preprocessing-split}, \ref{section-data_preprocessing-adjust}), feeding the retinal model with images (Step \ref{section-data_preprocessing-feed}), extracting and processing RGC responses (Steps \ref{section-data_preprocessing-valid}, \ref{section-data_preprocessing-combine}, \ref{section-data_preprocessing-concatenate}) and finally, feeding them to the classifiers. All steps, together with the corresponding design decisions, are analyzed below.

\subsubsection{Input} \label{section-data_preprocessing-input}
Transform each image from rgb to grayscale, only for CIFAR-10 and Imagenette. Then, normalize images in range (0, 1) and finally, reshape each image to (50, 50, 1) to match the input size of the retinal model.

\subsubsection{Data augmentation} \label{section-data_preprocessing-data_augmentation}
Implement data augmentation as a standard process to improve model training. Two of the following transformations are applied to each image: \textit{rotate} $45^{o}$ or $-45^{o}$, use \textit{gaussian noise} of scale = $0.1 * 255$ or scale = $0.2 * 255$, \textit{crop} images by 5 or 7 pixels from each side, \textit{translate} images over $x$ or $y$ axis by a percentage ranging from $-10\%$ to $+10\%$ of the image size. Parameter values are selected, so as to produce different perspectives of the same image and, at the same time, preserve the object for detection inside the frame of the image. Data augmentation is implemented using \textit{imaug} library\footnote{https://github.com/aleju/imgaug}.

% . We choose two different transformations among rotation, gaussian noise, crop and translation across $x$ and $y$ axis. More specifically, we rotate images by $45^{o}$ or $-45^{o}$ in order to model rotations that can be found in real life. Given the low resolution of the available images, we use gaussian noise of scale = $0.1 * 255$ or scale = $0.2 * 255$ in order to preserve the semantics of the images. We also crop images only by 5 or 7 pixels from each side in order to keep the main parts of the object within the frame. For the same reason, we translate images only by a percentage ranging from $-10\%$ to $+10\%$ of the image size. Data augmentation is implemented using imaug library \cite{imgaug}. After data augmentation, every image in the dataset has been replaced by three other images: the initial image, the one transformed using the first transformation and the one transformed using the second transformation. An example of data augmentation is shown in Fig. \ref{data-augmentation-example}.

% \begin{figure}[htbp]
% \centerline{\includegraphics[width=5cm]{data-augmentation-example.png}}
% \caption{An example of data augmentation.}
% \label{data-augmentation-example}
% \end{figure}

% \subsubsection{NORMALIZATION} Normalize images, such that every pixel has value between (0, 1).

\subsubsection{Split} \label{section-data_preprocessing-split} 
Split image in $p^{2}$ parts, where $p = 1, 2, 3, 4$ and reshape the size of each part to (50, 50). For $p=1$, we assume that the image is not split at all. Motivation behind image splitting is to artificially increase the limited number of output neurons ($60$ in our case) provided by the retinal model, so as to increase the performance of the classifiers.

\subsubsection{Adjust} \label{section-data_preprocessing-adjust} 
The goal of this step is to create a temporal dataset, as described in Section \ref{section-temporal-dataset}. It should be decided whether to repeat each image five times and get the full response trend for t = 0 ... 50 ms (\textit{Adjust=yes}) or have one row per image with ten image repetitions (see red frames in Fig. \ref{tempDataset-withRedLines}) and get a snapshot of the response at t = 100 ms (\textit{Adjust=no})\footnote{We found that the retina response at t=100 ms has high variance and thus, it could better distinguish different objects (data not shown).}.

% The goal of this step is to create a temporal dataset (Section \ref{section-temporal-dataset}). We should decide if we will adjust the dataset by repeating every image five times (Fig. \ref{tempDataset-withRedLines}). In case we do not implement adjustment, the temporal dataset will consist only of one row per image, with 10 image repetitions (i.e. 100 ms) in each row (see red frames in Fig. \ref{tempDataset-withRedLines}). 

\subsubsection{Feed} \label{section-data_preprocessing-feed}
The retinal model, described in Section \ref{section-retinal_model}, is fed with images and predicts retinal responses for 60 neurons.

\subsubsection{Valid} \label{section-data_preprocessing-valid}
In this step, the receptive fields of the neurons are identified using STA Analysis \cite{Chichilnisky2001WhiteNoise}. If STA manages to compute the center of the receptive field for a neuron, then we consider this neuron as valid. For the retinal model used, only 12 valid neurons are found. Therefore, it should be decided whether to keep all 60 neurons (\textit{Valid=no}) or only the 12 valid ones (\textit{Valid=yes}).

\subsubsection{Combine} \label{section-data_preprocessing-combine}
If adjustment was previously implemented, there are five arrays of RGC responses for every image, which are combined in one array, by applying elementwise \textit{min} or \textit{max} transformation.

\subsubsection{Concatenate} \label{section-data_preprocessing-concatenate}
If image splitting is implemented in Step \ref{section-data_preprocessing-split}, there is one array of 60 RGC responses for every part of the split image. These arrays are then concatenated to one larger array of size $a * (p^{2})$, where $a = 60$ (or $a = 12$ if valid neurons are selected) and $p = 2, 3, 4$. The set of final arrays is the training dataset for the classifiers and thus, the size of final arrays represents the number of features for the classifiers.

% \begin{figure*}[htbp]
% \centerline{\includegraphics[width=15cm, height=15cm]{wholepipeline.png}}
% \caption{An example illustrating basic preprocessing steps (Section \ref{section-data-preprocessing-steps}) for simulation A2 (Table \ref{types-of-simulations}). (I) Initial image is split in four parts (\textit{Split=yes, p = 2}). (II) We then create the temporal dataset by repeating each part five times (\textit{Adjust=yes}). (III) For each row of the temporal dataset, retinal model predicts RGC responses for 60 neurons and we choose all of them (\textit{Valid=no}). (IV) Every five rows, arrays of RGC responses correspond to different parts of the split image. We combine these arrays into one array of size 60 for each part, using elementwise max transformation (\textit{Combine=max}). (V) In the end, we concatenate all arrays of the four split parts and create one larger array of size 240, which is the input array to the classifiers.}
% \label{wholepipeline}
% \end{figure*}

\subsection{Retinal Model} \label{section-retinal_model}

We use a 3-layer CNN retinal model \cite{McIntosh2016} that was trained to predict response rates for sixty simultaneously recorded RGCs. This model was chosen, as it can effectively predict retinal responses to natural images and, being trained with natural images, it can model a wide range of retina’s biological properties. To train the retinal model, we used an image dataset consisting of $4890$ grayscale natural images of size $50$x$50$ pixels and the recorded retinal responses (retinal responses were recorded at Prof. E. Fernandez lab) \cite{lozano2018}. Each frame, corresponding to $10$ ms of visual stimulus, was projected onto the retina of a mouse for a total of $50$ ms. Thus, each frame was repeated five times and the whole dataset consists of $24450$ natural images of total duration $244.5$ s. 

% \begin{figure}[htbp]
% \centerline{\includegraphics[width=4cm]{natural-images-sample.png}}
% \caption{Sample images from the dataset used in retinal model’s training \cite{lozano2018}.}
% \label{natural-images-sample}
% \end{figure}

\subsection{Temporal dataset} \label{section-temporal-dataset}

\begin{figure}[htbp]
\centerline{\includegraphics[width=6.6cm]{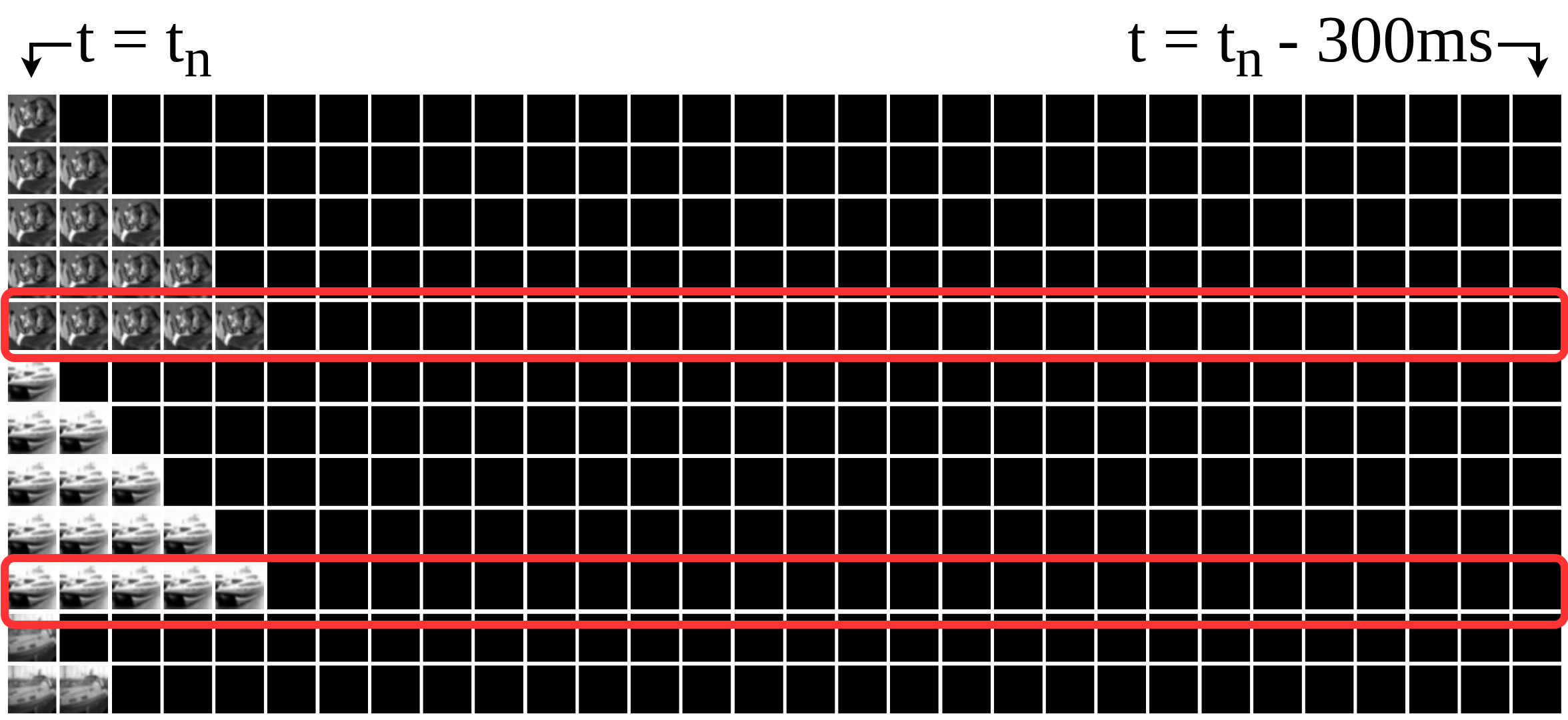}}
\caption{A clipping of the temporal dataset. Every row in the temporal dataset represents an input sample to the retinal model and it has \textit{temporal\_interval} frames (here, \textit{temporal\_interval} $=30$). In each row, the number of image repetitions ($n = 1, 2, 3, 4, 5$) represents the time that the image has been projected onto the retina ($\Delta t = 10, 20, 30, 40, 50$ ms respectively). Black frames represent the intermediate stage between the alternation of images, where no light is projected onto the retina.}
\label{tempDataset-withRedLines}
\end{figure}

The response of RGCs depends not only on the current stimulus, but also on preceding stimulations. To model this temporal dependency, a temporal dataset is created. For each image being projected onto the retina, we keep track of the history of images projected before. The total number of frames used -the actual image projected onto the retina (at $t = t_n$) plus the additional image frames accounting for the stimulus history- is called \textit{temporal\_interval}. Every row in the temporal dataset
represents an input sample to the retinal model. The number of image repetitions ($n$) represents the duration ($n*10$ ms, given that each frame corresponds to 10 ms of visual stimulus) that the retina is exposed to a specific image (Fig. \ref{tempDataset-withRedLines}).

% \begin{figure}[thpb]
%   \centering
%   \framebox{\parbox{3in}{\includegraphics[width=\linewidth]{tempDataset-usedinSimulations.png}
%  }}
%   %\includegraphics[scale=1.0]{tempDataset-usedinSimulations.png}
%   \caption{Inductance of oscillation winding on amorphous
%   magnetic core versus DC bias magnetic field}
%   \label{figurelabel}
% \end{figure}

% \begin{figure}[htbp]
% \centerline{\includegraphics[width=\linewidth]{tempDataset-noAdjust.png}}
% \caption{ A clipping of a temporal dataset, where no adjustment is implemented. There is only one row per image, with 10 image repetitions in each row.}
% \label{tempDataset-noAdjust}
% \end{figure}

\subsection{Design of simulations} \label{section-design_of_simulations}

We conducted simulations for all different combinations of design decisions during data preprocessing. In particular, we trained four different classifiers: MLP\_500\_100 (size of input layer = 500, size of hidden layer = 100), MLP\_n\_n/2 (size of input layer = n, size of hidden layer = n/2, where n = number of features), SVM (kernel=‘rbf’) and Random Forest (max\_features=12). Hyperparameters were chosen based on the performance of classifiers in initial simulations. We also chose four datasets (Table \ref{table-datasets}): CIFAR-10, MNIST, Fashion MNIST and Imagenette with ten different classes (ten object categories) each, so as to compare classification tasks with equal number of classes. Then, 10000 samples from each dataset were augmented, creating in this way 30000 samples, which were randomly divided according to 70\%/30\% train/test split. For each simulation, we trained each classifier ten times and we calculated the mean and standard deviation of accuracy. We repeated all simulations for two different parametrizations of the CNN retinal model (Section \ref{section-retinal_model}): \textit{RetModel1} with \textit{temporal\_interval}=30 and \textit{RetModel2} with \textit{temporal\_interval}=40.

\begin{table}[h]
\caption{Description of the datasets used in simulations.}
\label{table-datasets}
\begin{center}
\begin{tabular}{|c|c|c|c|c|}
\hline
\textbf{Dataset} & \textbf{Image} & \textbf{Type of} & \textbf{\# Classes}\\
\textbf{} & \textbf{size} & \textbf{ classification} & \textbf{}\\
\hline
CIFAR-10 \cite{cifar10-Krizhevsky09learningmultiple} & 32x32 & objects & 10\\
\hline
MNIST \cite{lecun2010mnist} & 28x28 &  digits & 10\\
\hline
Fashion MNIST \cite{fashion_mnist} & 28x28 & clothing items & 10\\
\hline
Imagenette \tablefootnote{Available at https://github.com/fastai/imagenette/} & variable & objects & 10\\
\hline
\end{tabular}
\end{center}
\end{table}

% \begin{table}[htbp]
% \caption{Five types of simulations for different preprocessing decisions}
% \begin{center}
% \begin{tabular}{|c|c|c|c|c|}
% \hline
% & \textbf{Combine} & \textbf{Augment} & \textbf{Adjust} & \textbf{Valid} \\
% \hline
% \textbf{A1} & yes & min & yes & no  \\
% \hline
% \textbf{A2} & yes & max & yes & no  \\
% \hline
% \textbf{B1} & yes & max & no & yes  \\
% \hline
% \textbf{C1} & yes & max & no & no   \\
% \hline
% \textbf{D1} & yes & max & yes & yes \\
% \hline
% \end{tabular}
% \label{types-of-simulations}
% \end{center}
% \end{table}

\section{Results and Discussion}

\subsection{Preprocessing decisions for functional assessment} \label{results-design_decisions}

\begin{figure}[htbp]
\centerline{\includegraphics[height=11.2cm]{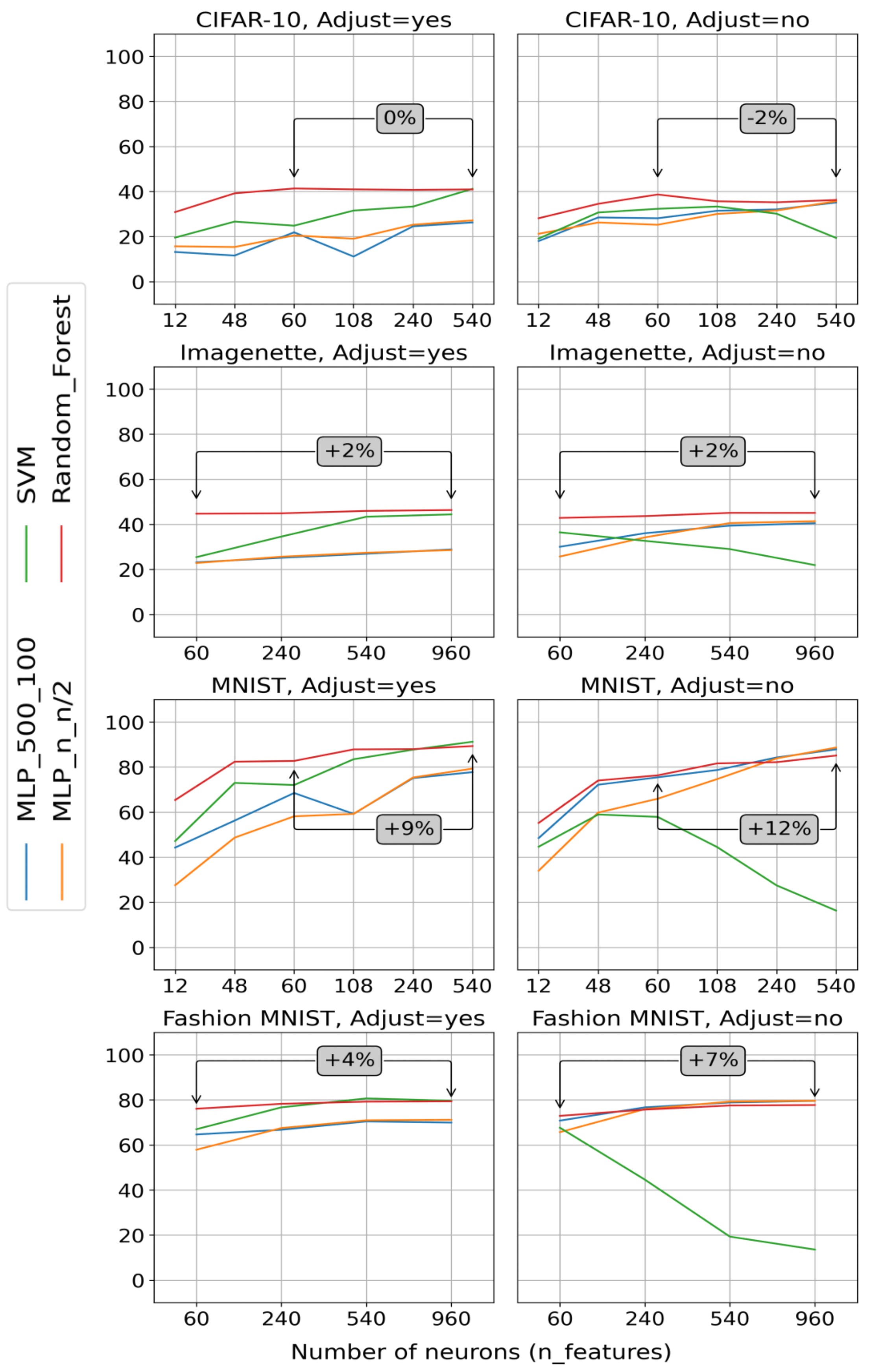}}
\caption{Sensitivity plots correlating the accuracy of classifiers (\mbox{CIFAR-10}, MNIST, Fashion MNIST, Imagenette) with the number of neurons used as input (\textit{n\_features}), for both \textit{Adjust=yes/no} (Step \ref{section-data_preprocessing-adjust}).}
\label{sensitivity-plots}
\end{figure}

Initially, we performed simulations with CIFAR-10 and MNIST, splitting images in \textit{None}, $4, 9$ parts ($p=1,2,3$ respectively) for both \textit{Valid=yes/no} (Step \ref{section-data_preprocessing-valid}). Then, we split images of the two more complex datasets, Fashion MNIST and Imagenette, in \textit{None}, $4, 9, 16$ parts for only \textit{Valid=no}. In \hbox{Fig. \ref{sensitivity-plots}}, we see the percentage differences between 60 neurons (\textit{Split=None}) and max splitting (\textit{Split=9} or \textit{Split=16}). We observe that image splitting (i.e. using more than 60 neurons) increases the performance mostly in MNIST (>10\%) and less in Fashion MNIST (<10\%), while it does not further improve performance in \hbox{CIFAR-10} and Imagenette. This can be explained by structural differences mentioned in Section \ref{results-datasets}. We further assume that, by splitting images in too many parts, the objects are difficult to be recognized due to oversegmentation and so, performance does not significantly improve with splitting. In addition, if we compare plots with \textit{Adjust=yes} and those with \textit{Adjust=no} (Fig. \ref{sensitivity-plots}), we see that, keeping a snapshot of the retina response at a critical $t$ (at which retina response has high variance), produces similar results as keeping the full trend of the retina response over time. By keeping only the critical responses, we can also save significant computational resources and time. Finally, we tested a set of different classifiers and we found that the type of classifier is not an important  factor -our conclusions remain unchanged, irrespective of the classifier we use- even if the Random Forest had the most efficient and consistent performance across the simulations (Fig. \ref{sensitivity-plots}).

%Results also highlighted the need for prosthetic devices to be equipped with a sufficiently large number of electrodes. More electrodes can stimulate a larger number of RGCs and provide higher resolution vision;  enabling implantees to solve more complex visual tasks \cite{fernandez2018}.

% Finally, we examined which classifiers perform better in such types of tasks and  Random Forest was proved to be the best performing and the most consistent classifier.

% Even if SVM sometimes approached or even exceeded Random Forest’s accuracy, it did not perform that well across the whole range of simulations.

\subsection{Functional assessment performance on different datasets}\label{results-datasets}

\begin{figure}[htbp]
\centerline{\includegraphics[width=7cm, height=4.8cm]{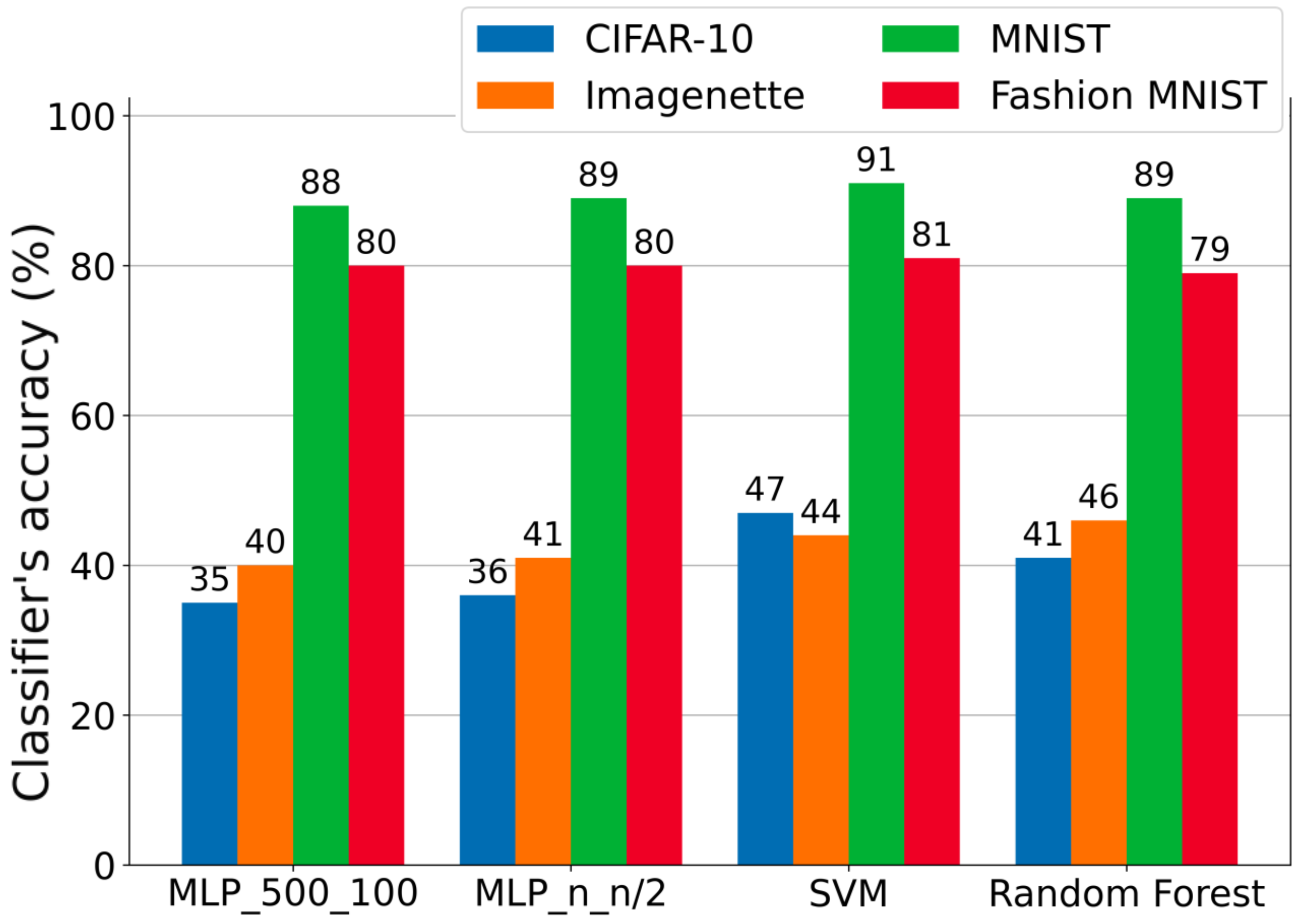}}
\caption{Comparison of the maximum performance of classifiers between datasets.}
\label{max_accuracy_for_datasets}
\end{figure}

Fig. \ref{max_accuracy_for_datasets} compares the maximum performance of classifiers between different visual understanding tasks, i.e. different datasets (Table \ref{table-datasets}). CIFAR-10 and Imagenette achieve $\sim$40\% accuracy, while MNIST and Fashion MNIST exceed 80\%. We see that image resolution is not a crucial factor in the model’s performance. Although Imagenette was compared to CIFAR-10 to test images with higher resolution, it performs only slightly better -and in some cases- worse than \hbox{CIFAR-10}. Furthermore, MNIST and Fashion MNIST have twice the accuracy of CIFAR-10 and Imagenette, even if they consist of images with lower resolution. Significant differences in performance between MNIST/Fashion MNIST and CIFAR-10/Imagenette can be explained, if we take a closer look at the structural differences between these datasets. Images in both MNIST and Fashion MNIST have a dark background with the object for classification situated in the middle. On the other hand, CIFAR-10 and Imagenette consist of real-life images, with unclear background-foreground segregation and more complex structures. This plays a key role, if we additionally consider the  retina only reacts in spatiotemporal changes among different image frames \cite{The_New_Visual_Neurosciences}. This retina's property favors both MNIST and Fashion MNIST, where only the object under classification changes over different dataset images.

% \begin{figure}[htbp]
% \centerline{\includegraphics[width=6cm]{mnist_vs_fashionMnist.png}}
% \caption{Comparison of the maximum performance of classifiers between MNIST and Fashion MNIST.}
% \label{mnist_vs_fashionMnist}
% \end{figure}

% Classifiers could be attached to the previous subsection \subsection{Classifiers}

\subsection{Functional assessment of different retinal models}

Finally, functional assessment is applied in order to compare the performance of two different retinal models, \mbox{\textit{RetModel1}} and \textit{RetModel2} (Section \ref{section-design_of_simulations}). Results show that the retinal model with the lowest mean squared error (MSE) (i.e. the one being closer to the biological retina) performs better across the whole range of functional simulations. Therefore, it seems that functional assessment is in accordance with standard evaluation techniques. Furthermore, by functionally assessing a model, we get a direct and more easily interpretable estimate of how well an implantee may perform in a visual task of interest.

\subsection{Limitations and future directions of functional assessment}

During the development of this work, several limitations were encountered. We used a retinal model that has been trained to predict only a limited number of RGC responses (Section \ref{section-retinal_model}). However, a higher number of RGC responses could better represent complex images and provide classifiers with a richer set of features to solve more efficiently real-life visual tasks. Therefore, there is a need to increase the number of RGC responses given by retinal models, either by using improved experimental methods to collect the data  or using a different architecture for the retinal model. 

Another limitation arises from choosing to process retinal responses, which are one-dimensional firing-rate arrays, with traditional machine learning classifiers. From a biological perspective, the visual pathway includes neural processing both in the retina and mainly, in the brain’s visual centers, which are responsible for higher visual functionalities, like object recognition. Literature indicates that the brain's visual centers can be effectively modeled by deep neural architectures \cite{Schrimpf_BrainScore_2020}. Deep learning networks have been also used, in high-impact research on biological vision,  to model the ventral visual stream in order to elucidate retinal mechanisms \cite{Lindsey2019}. Taking those insights into consideration, we suggest that future efforts for FA should be focused on combining end-to-end deep learning architectures to model both the retina and the rest of the visual pathway. Moreover, we may train on tasks with unsupervised methods, which produce biologically plausible ventral visual system models and follow bio-plausible sensory learning procedures \cite{Zhuang2021}. Given also the interpretable nature of FA, future development involves explainable models, providing insights into the relationship between image properties and retina output \cite{Athanasiou2020}.

% The need for increased RGC responses has also implications on retinal prostheses, where more electrodes are required to respond to complex visual stimuli; thus, restoring implantees a useful vision \cite{fernandez2018}.

% also highlights the need for future prosthetic devices to be equipped with a sufficiently large number of electrodes, so as to stimulate a larger number of RGCs;  enabling implantees to solve more complex visual tasks \cite{fernandez2018}.

%Results also highlighted the need for prosthetic devices to be equipped with a sufficiently large number of electrodes. More electrodes can stimulate a larger number of RGCs and provide higher resolution vision;  enabling implantees to solve more complex visual tasks \cite{fernandez2018}. 

\section{Conclusion}

In this work, we introduced the concept of functional assessment and designed a machine learning framework for it. We investigated how retinal models, trained to faithfully reproduce retina output, perform in visual understanding tasks.  We show that FA is comparable with the established evaluation method; yet, FA provides a direct and easily interpretable way of assessment, based on the performance on visual tasks. We also found that performance in FA is closely dependent on the given task. Finally, image splitting, as a way to increase the number of output neurons, does not significantly improve accuracy; still, restoring functional vision requires that retinal models interpret images to a larger number of RGC responses.

% Finally, image splitting, as a way to increase the number of output neurons, does not significantly improve accuracy; still, a larger number of RGC responses provided by the retinal model is needed to solve real-life visual tasks.

\bibliographystyle{IEEEtran}
\bibliography{references}

\addtolength{\textheight}{-12cm}   % This command serves to balance the column lengths

\end{document}